\documentclass[journal,10pt]{IEEEtran} 
\usepackage[figurename=Fig.]{caption}
\usepackage{tikz}
\usetikzlibrary{calc, positioning, shapes.geometric, decorations.pathreplacing}
\usepackage{array}
\usepackage{verbatim}
\usetikzlibrary{calc, shapes, arrows, positioning}
\usepackage{authblk}
\usepackage{blindtext}
\usepackage{ifpdf}
\usepackage{graphicx}
\usepackage{tabulary}
\usepackage{amsmath,amsthm,amssymb}
\usepackage{amsmath}
\usepackage{kantlipsum}
\allowdisplaybreaks
\usepackage{cleveref}
\usepackage{lipsum}%
\usepackage{optidef}
\usepackage[english]{babel} 
\theoremstyle{plain}

\crefname{theorem}{Theorem}{theorem}
\crefname{lemma}{Lemma}{Lemmas}

\usepackage{cite}
\usepackage{dsfont}
\usepackage[justification=centering]{caption}
\usepackage{color}
\usepackage{algorithm}
\usepackage{algorithmicx, algpseudocode}
\usepackage{etoolbox}
\usepackage{subcaption}
\usepackage{balance}
\usepackage{empheq,etoolbox}
\usepackage[utf8]{inputenc}
\usepackage{multirow}
\usepackage{float}

\usepackage{amssymb}% http://ctan.org/pkg/amssymb
\usepackage{pifont}% http://ctan.org/pkg/pifont
\usepackage[hmargin=1.27cm,top=1.3cm,bottom=1.3cm]{geometry}

\usepackage{babel}
\floatstyle{plaintop}
\restylefloat{table}
\patchcmd{\subequations}
\usepackage{lipsum} % For some dummy text
\allowdisplaybreaks

\tikzset{brace/.style={decorate, decoration={brace}},
 brace mirrored/.style={decorate, decoration={brace,mirror}},
}

\newcounter{brace}
\newcounter{arrow}
\begin{document}
 \captionsetup[figure]{name={Fig.},labelsep=period}

\title{Recent Advances in Near-Field Beam Training and Channel Estimation for XL-MIMO Systems}

\bstctlcite{IEEEexample:BSTcontrol}
\author{Ming Zeng, Ji Wang, Wanming Hao, Zheng Chu, Wenwu Xie and Quoc-Viet Pham
    \thanks{M. Zeng is with the Department of Electric and Computer Engineering, Laval University, Quebec City, Canada (email: ming.zeng@gel.ulaval.ca).}
    
    \thanks{J. Wang is with the Department of Electronics
and Information Engineering, Central China Normal University, Wuhan, China (e-mail: jiwang@ccnu.edu.cn).}

%\thanks{X. Li is with the School of Physics and Electronic Information Engineering, Henan Polytechnic University, Jiaozuo, China (email: lixingwang@hpu.edu.cn).}

\thanks{W. Hao is with the School of Electrical and Information Engineering, Zhengzhou University, Zhengzhou, China (e-mail: wmhao@hotmail.com).}

\thanks{Z. Chu is with the Department of Electrical and Electronic Engineering, University of Nottingham, Ningbo, China (Email: andrew.chuzheng7@gmail.com).}

\thanks{W. Xie is with the School of Information Science and Engineering, Hunan Institute of Science and Technology, Yueyang, China (e-mail: gavinxie@hnist.edu.cn).}

\thanks{Q.-V. Pham is with School of Computer Science and Statistics, The University of Dublin, Dublin, Ireland (e-mail: Viet.Pham@tcd.ie).}

    }
\maketitle

\begin{abstract}
Extremely large-scale multiple-input multiple-output (XL-MIMO) is a key technology for next-generation wireless communication systems. By deploying significantly more antennas than conventional massive MIMO systems, XL-MIMO promises substantial improvements in spectral efficiency. However, due to the drastically increased array size, the conventional planar wave channel model is no longer accurate, necessitating a transition to a near-field spherical wave model. This shift challenges traditional beam training and channel estimation methods, which were designed for planar wave propagation. In this article, we present a comprehensive review of state-of-the-art beam training and channel estimation techniques for XL-MIMO systems. We analyze the fundamental principles, key methodologies, and recent advancements in this area, highlighting their respective strengths and limitations in addressing the challenges posed by the near-field propagation environment. Furthermore, we explore open research challenges that remain unresolved to provide valuable insights for researchers and engineers working toward the development of next-generation XL-MIMO communication systems.

\end{abstract}

\IEEEpeerreviewmaketitle

\section{Introduction}
\label{Sec:Introduction}
With the commercial rollout of fifth-generation (5G) networks, both academia and industry have been actively exploring key technologies for the sixth-generation (6G) wireless systems. Among these, extremely large-scale multiple-input multiple-output (XL-MIMO) has emerged as a promising candidate \cite{Wang_WCM23, Lu_COM24, Liu_COMMST25, Lei_WCM, Yang_TSPM25, Abdallah_TSPM, Wang_TWC24}. By equipping base stations (BSs) with an order of magnitude more antennas than conventional massive MIMO, XL-MIMO enables significantly higher spatial resolution and spectral efficiency.
However, the larger array size, coupled with shrinking cell coverage in modern communication networks, increases the likelihood of user equipments operating in the near-field region. In this scenario, the conventional planar wave assumption becomes inaccurate, necessitating a shift to a more precise near-field spherical wave model. 
Unfortunately, the shift from planar to spherical wave propagation invalidates many fundamental techniques used in current MIMO systems, including beam training, channel estimation, and beamforming. In particular, the near-field spherical wave model poses significant challenges for beam training and channel estimation, which are critical for realizing the promised gains in spectral efficiency. Thus, investigating their impact and developing effective solutions is essential for the successful deployment of XL-MIMO.

Beam training is a fundamental technique in 5G systems, facilitating the establishment of high signal-to-noise ratio (SNR) links for data transmission. In this process, the BS transmits signals across multiple beam directions, enabling the user equipment to identify the optimal beam with the strongest signal strength. This alignment is especially critical in high-frequency bands channels, such as millimeter wave (mmWave) and terahertz (THz), where precise beamforming is essential for reliable communication. 
In conventional far-field communication systems, the discrete Fourier transform (DFT)-based codebook is widely used to help user equipment identify the optimal transmission angle. However, in spherical wave channel models, the phases of multipath components vary nonlinearly with the antenna index due to their coupling with distance. As a result, directly applying the DFT codebook leads to an energy-spread effect, making it difficult to accurately determine the best beam. This challenge necessitates novel codebook designs that consider both angle and distance. Furthermore, integrating distance into beam training allows for the differentiation of users in similar angular directions, enhancing beam focusing and improving spatial resolution.

The channel information obtained from beam training is inherently limited, as only the signal strength is fed back. To fully exploit the potential of XL-MIMO and enhance multi-user communications, comprehensive channel estimation is necessary to acquire the complete channel state information (CSI). Conventional channel estimation techniques employed in massive MIMO, such as minimum mean square error (MMSE) and least squares (LS), can be directly extended to time division duplex (TDD)-based XL-MIMO systems with fully digital precoding. However, for TDD systems with hybrid precoding—commonly adopted in XL-MIMO to minimize the number of radio-frequency (RF) chains—and frequency division duplex (FDD) systems, significant challenges arise when applying existing compressive sensing (CS)-based algorithms developed for conventional massive MIMO. These methods typically exploit channel sparsity\footnote{{\color{black}To avoid potential ambiguity, it should be emphasized that the term channel sparsity in the context of near-field XL-MIMO refers to sparsity in the parameter domain rather than structural sparsity of the channel matrix itself.}} in the angular domain to reduce pilot overhead, an assumption that no longer holds under the near-field spherical wave propagation model. Hence, novel channel estimation algorithms that jointly leverage sparsity in both the angular and distance domains are essential for effective CSI acquisition in near-field XL-MIMO systems.

Motivated by the aforementioned discussions, this article provides a comprehensive evaluation of state-of-the-art advancements in beam training and channel estimation for XL-MIMO systems while identifying key open challenges. The primary contributions of this work are summarized as follows:

\begin{itemize}
    \item {\bf{Structured Survey}}: We present a systematic and categorized survey of beam training and channel estimation techniques specifically designed for XL-MIMO systems, offering a clear taxonomy of existing approaches.

    \item {\bf{In-Depth Analysis}}: We examine the fundamental principles, core methodologies, and recent developments in this domain, emphasizing their strengths and limitations in addressing the unique challenges posed by near-field propagation.
    
    \item {\bf{Identification of Open Challenges}}: We highlight critical challenges and unresolved issues that require further research, providing insights into potential directions for future advancements in XL-MIMO beam training and channel estimation.
\end{itemize}

{\color{black}
While several recent surveys have explored near-field MIMO systems \cite{Wang_WCM23, Lu_COM24, Liu_COMMST25, Lei_WCM, Yang_TSPM25, Abdallah_TSPM}, our work is distinguished by its focused scope and systematic depth. Unlike \cite{Wang_WCM23, Lu_COM24, Liu_COMMST25}, which provide broad overviews of near-field MIMO across multiple domains, this article concentrates specifically on the core challenges of beam training and channel estimation---two essential pillars for unlocking the potential of XL-MIMO in near-field environments. Compared to the more concise tutorial-style magazine papers \cite{Lei_WCM, Yang_TSPM25, Abdallah_TSPM}, our survey offers a more structured and comprehensive taxonomy of state-of-the-art techniques tailored to these two problems. In particular, \cite{Lei_WCM} and \cite{Yang_TSPM25} focus mainly on channel estimation for localization applications, while \cite{Abdallah_TSPM} emphasizes the role of the effective beam-focusing Fraunhofer distance (EBFD) in beam training. By contrast, our work not only synthesizes recent advances in both beam training and channel estimation but also highlights open research directions, offering a holistic and in-depth resource for advancing near-field XL-MIMO research.
}

{\color{black}
This article is organized as follows: Section II introduces the fundamentals of XL-MIMO. Sections III and IV present the key techniques for beam training and channel estimation in XL-MIMO systems, respectively. Section V highlights potential challenges and open directions, while Section VI draws the conclusion. 
}

\begin{figure}
\centerline{\includegraphics[width=3.5in]{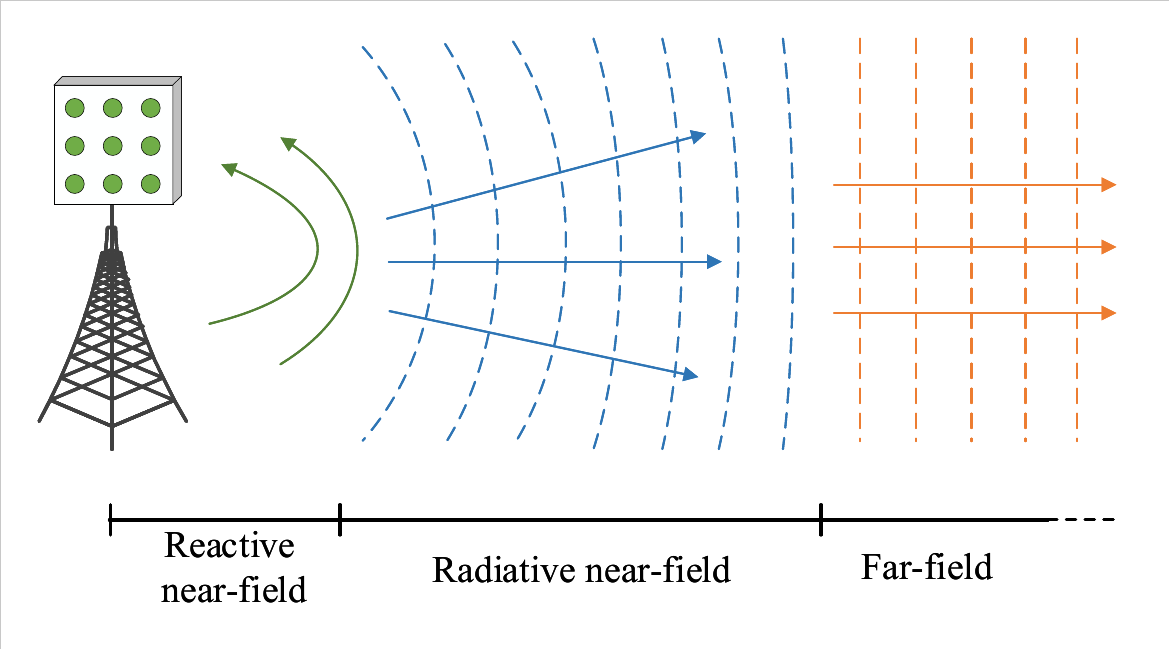}}
\caption{Spherical wave illustration of near-field channel.\label{fig1}}
\end{figure}

\begin{figure}
\centerline{\includegraphics[width=3.5in]{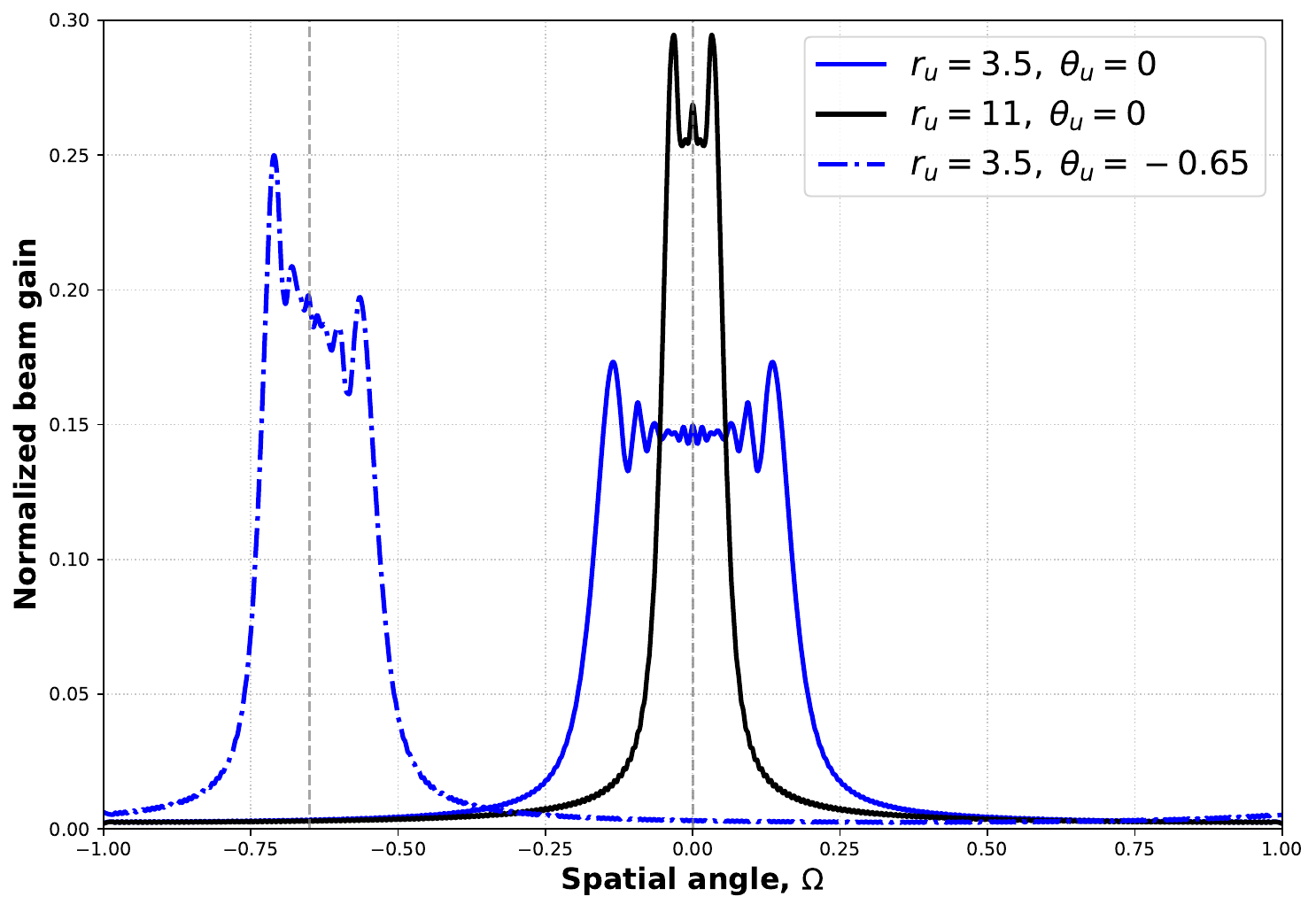}}
\caption{Beam energy-spread effect (received beam pattern under far-field beamformers for different user angles and ranges under 256 antennas at 30 GHz).\label{fig2}}
\end{figure}

\section{Fundamentals of XL-MIMO Systems}
\label{Sec:LC}
XL-MIMO has garnered significant attention as a key enabler for 6G and beyond networks. By deploying an order of magnitude more antennas than conventional massive MIMO at the BSs, XL-MIMO not only enhances multi-user connectivity but also substantially improves spectral and energy efficiency. Notably, studies have demonstrated that XL-MIMO can achieve a considerable capacity gain over existing MIMO systems. 
However, the dramatic increase in array size leads to an extended Rayleigh/Fraunhofer distance, a fundamental criterion for delineating the near-field and far-field regions. Concurrently, the continuous reduction in cell size in modern communication networks exacerbates this effect, making it increasingly unlikely for user equipment to remain within the conventional far-field region. In the near-field regime, the conventional planar wave assumption becomes invalid, necessitating the adoption of a more accurate spherical wave-based channel model (shown in Fig. 1 to capture the spatial characteristics of signal propagation. 

In a spherical wave-based channel model, the phase of the array response vector is no longer a linear function of the array element indices, as observed in the conventional planar wave assumption. Instead, it is intricately coupled with the element-wise propagation distance in a nonlinear manner. This nonlinear phase-distance relationship significantly alters the spatial characteristics of wavefront propagation in XL-MIMO systems. 
As illustrated in Fig. 2, this phenomenon gives rise to the so-called beam energy-spread effect when applying conventional DFT-based beamforming \cite{Cui_TCOMM_22}. Specifically, in the far-field regime, DFT beams are designed under the assumption of a linear phase progression across array elements, ensuring coherent beamforming gain at the intended direction. However, the phase evolution is nonlinear in the near-field region such that the direct application of DFT beams leads to energy dispersion over a wider angular span, reducing beamforming gain and impairing both beam training and channel estimation accuracy. This degradation poses a fundamental challenge in XL-MIMO system design.

\begin{figure}
\centerline{\includegraphics[width=3.5in]{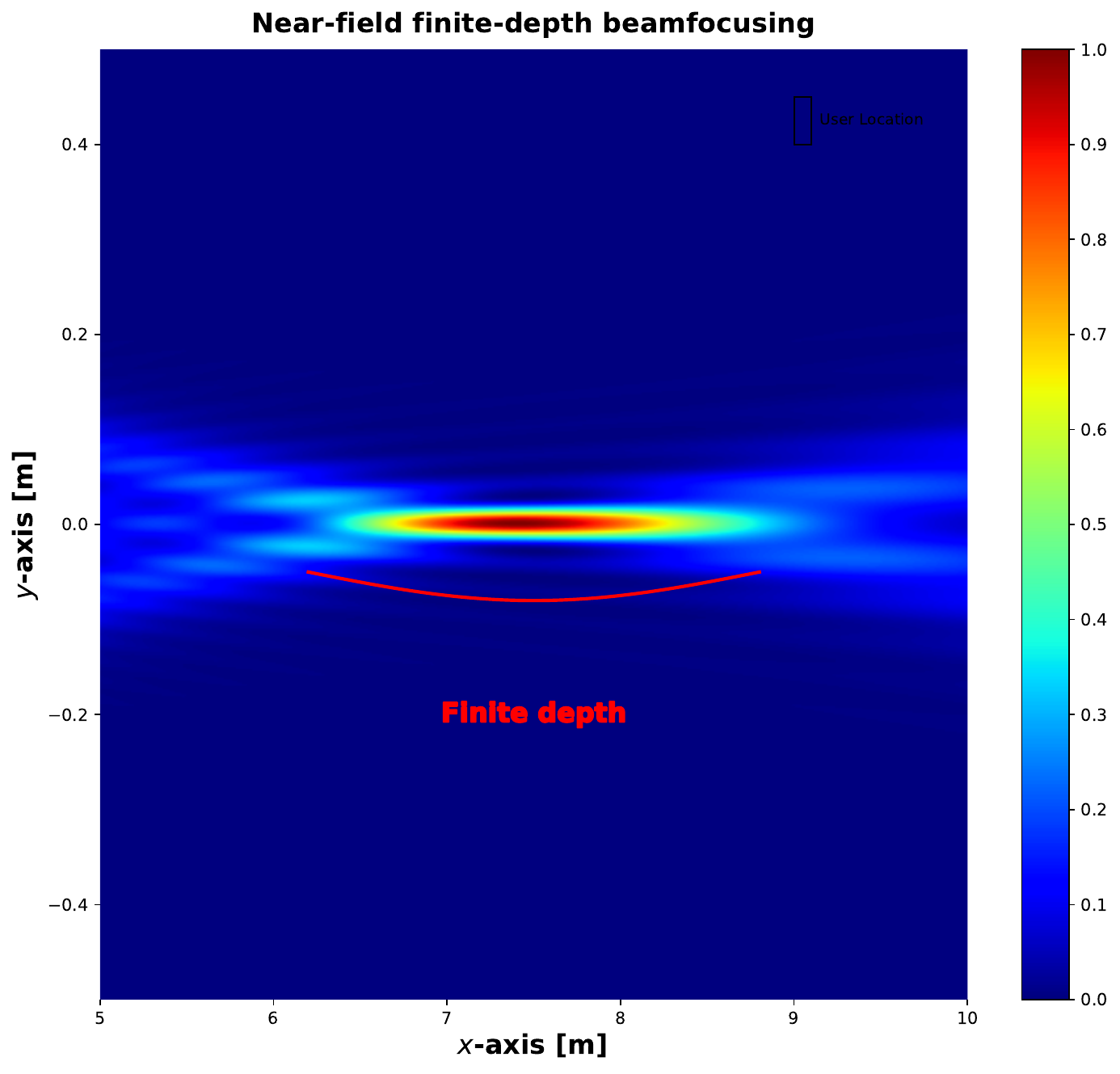}}
\caption{Finite depth of beamfocusing (512 antennas at 100 GHz).\label{fig3}}
\end{figure}

On the other hand, the inherent coupling of phase and distance in the array response vector enables finite-depth beam focusing, a unique property of near-field spherical wavefronts, as shown in Fig. 3. Unlike far-field beams, which form a plane wavefront extending infinitely, near-field beams can concentrate energy at a specific spatial location, not only in angular dimensions but also in range. This capability allows BSs to distinguish multiple users that share the same angular direction but are located at different distances—a functionality absent in conventional MIMO systems, where angular separation is the sole basis for spatial multiplexing. The ability to focus beams at precise locations holds great potential for improving user separation, enhancing spatial multiplexing gains, and enabling new applications such as location-aware communications and holographic radio.

%Introduce the beam energy spread and finite beam length. Draw the three subfigures. Also, show the relation between Rayleigh distance and frequency under given antenna dimension (tutorial paper). 

%Introduce the challenge in beam training and channel estimation and mention that this will be addressed in the following sections. 

\section{Near-Field Beam Training and Codebook Design}
\subsection{Motivation}
\label{SubSec:LC_motivation}
%To support the high data rate requirements of beyond 5G and 6G networks, high-frequency bands such as mmWave and terahertz (THz) spectra are expected to be deployed in hotspot scenarios. However, as the operating frequency increases, channel conditions deteriorate due to severe path loss, higher penetration loss, and reduced multipath reflections. Thus, mmWave and THz systems are often line-of-sight (LoS) dominant, for which beam training represents a low-cost but highly effective solution to establish initial communication between the BS and user equipments, without the need of explicit CSI. More exactly, the BS utilizes a predefined beamforming codebook, such as a DFT-based codebook for beam alignment. The BS sequentially transmits beamformed signals corresponding to different angular directions, while the user equipment measures the received signal strength and selects the beam with the highest gain for communication. 

To meet the high data rate requirements, high-frequency bands, such as mmWave and THz spectra, are expected to be deployed in hotspot scenarios. However, as the operating frequency increases, channel conditions deteriorate due to severe path loss, higher penetration loss, and reduced multipath reflections. Consequently, mmWave and THz communication systems are often line-of-sight (LoS) dominant, necessitating efficient beam management techniques. Beam training serves as a low-complexity yet effective approach to establish an initial communication link between the BS and user equipment without requiring explicit CSI. Specifically, the BS first predefines a beamforming codebook, often the DFT-based one, for beam alignment. Then, the BS sequentially transmits beamformed signals using codewords from the predefined codebook, while the user equipment measures the received signal strength and selects the codeword with the highest gain for communication.

\begin{figure}
\centerline{\includegraphics[width=3.5in]{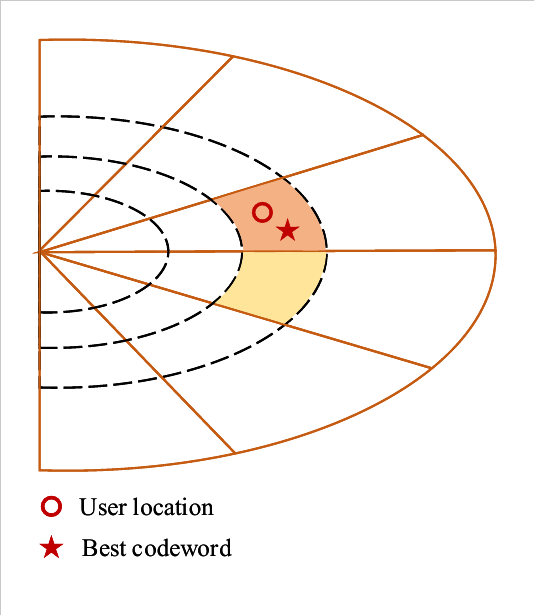}}
\caption{Polar-domain-based beam training \cite{Cui_TCOMM_22}.\label{fig4}}
\end{figure}

Although the DFT-based codebook is highly effective for beam training in far-field systems, it is not directly applicable to near-field scenarios. When a DFT-based codebook is used in the near-field, it results in beam energy dispersion, as illustrated in \text{Fig. 2}, which complicates the identification of the optimal beam angle. Additionally, applying such codewords for beamforming introduces significant power leakage to users with similar angular positions. Furthermore, the DFT-based codebook does not leverage the finite-depth beam focusing capability of the near-field, rendering it incapable of distinguishing users with identical angular coordinates but different radial distances. Therefore, for near-field XL-MIMO systems, novel beam training techniques and codebooks that jointly consider both angular and range dimensions must be developed.  

\begin{figure}
\centerline{\includegraphics[width=3.5in]{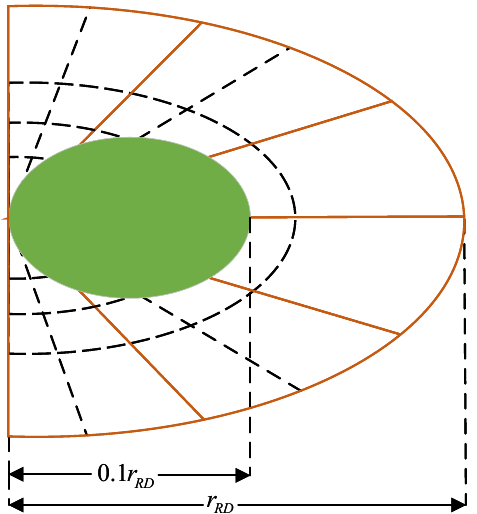}}
\caption{Effective Rayleigh distance \cite{Abdallah_TSPM, Hussain_WCNC24}.\label{fig5}}
\end{figure}

\subsection{State-of-the-art}
\label{SubSec:LC_state}
\subsubsection{Polar-domain-based beam training}
Polar-domain-based beam training has emerged as a promising approach for near-field XL-MIMO systems \cite{Cui_TCOMM_22}. As illustrated in Fig. 4, this method partitions the entire near-field region into discrete grids, each characterized by distinct angular and range coordinates. The codebook is then designed such that each codeword generates a beam that concentrates the majority of its energy within a single grid, effectively incorporating both angular and range information into the beam training process. Moreover, to minimize power leakage into adjacent beams, uniform sampling should be applied to the angular dimension, while non-uniform sampling is required for the range dimension to better capture the characteristics of near-field propagation. 

While polar-domain-based beam training effectively mitigates beam energy spread and enables finite-depth beam focusing for XL-MIMO systems, it significantly increases the number of codewords—from one dimensional angular search to two dimensional angular and range search. Consequently, this leads to substantially higher overhead during beam searching, thereby reducing the available time for data transmission. 

\begin{figure}
\centerline{\includegraphics[width=3.5in]{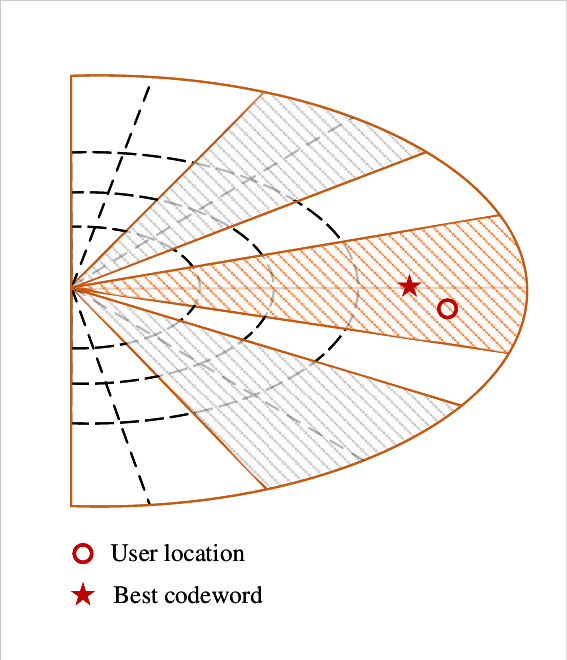}}
\caption{Beam training \cite{Weng_TVT_24}: a) far-field for angle estimation.}\label{fig6}
\end{figure}

%While polar-domain-based beam training effectively mitigates beam energy spread and enables finite-depth beam focusing for XL-MIMO systems, it encounters two key challenges. First, incorporating distance into the codebook design significantly increases the number of codewords—from one dimensional angular search to two dimensional angular and range search. Consequently, this leads to substantially higher overhead during beam searching, thereby reducing the available time for data transmission. Second, as an on-grid beam training method, it only optimally aligns with users whose angular and range coordinates coincide with predefined sampling points. However, since actual user positions are continuous and randomly distributed, any mismatch between the user’s true location and the nearest grid point results in beam training loss, degrading overall system performance.

Recently, several approaches have been proposed to reduce the overhead of on-grid polar-domain-based beam training \cite{Weng_TVT_24, zhou2024, Hussain_WCNC24}.
As illustrated in Figs 6 and 7, hierarchical codebooks have been introduced to mitigate the need for a two-dimensional exhaustive search. These codebooks first perform coarse angular alignment and then refine both angle and range estimation. For instance, the authors in \cite{Weng_TVT_24} propose a two-stage training protocol that separately handles angle and distance estimation in a uniform planar array-based system. In the first stage, a far-field codebook is used to obtain an approximate user angle. To address beam energy spread and reduce the search space in the angular domain, a wide codeword is constructed by summing multiple phase-shifted narrow far-field codewords. Once the optimal wide codeword is identified, an exhaustive search is conducted over the corresponding narrow codewords and range samples to jointly determine the user’s precise angle and range. This method achieves substantial reduction in training overhead compared to conventional beam sweeping.
Additionally, a sparse linear array configuration can be employed to simultaneously generate multiple beams by leveraging near-field grating lobes. In this approach, only a subset of antennas is activated, forming a sparse array that produces a primary main lobe alongside several grating lobes \cite{zhou2024}. As depicted in Figs 8 and 9, the corresponding multi-beam training procedure consists of two stages. First, multi-beam sweeping is performed over time to identify the best candidate beam. Next, single-beam sweeping is conducted using the entire antenna array to resolve ambiguities and pinpoint the user’s precise location.
Another effective strategy to reduce training overhead is to redefine the near-field region \cite{ Hussain_WCNC24}. The study in \cite{Abdallah_TSPM, Hussain_WCNC24} argues that the classical Rayleigh distance overestimates the near-field range when assessing channel capacity. Instead, the effective Rayleigh distance, which accounts for beamforming gain loss when using far-field beamforming vectors, provides a more accurate characterization of the near-field region. Under a 3 dB beamforming gain loss criterion, the effective Rayleigh distance is approximately ten times shorter than the classical definition, as shown in Fig. 5. By adopting this refined metric, the authors in \cite{ Abdallah_TSPM, Hussain_WCNC24} achieve a $50\%$ reduction in codebook size compared to \cite{Cui_TCOMM_22}. 

\begin{figure}
\centerline{\includegraphics[width=3.5in]{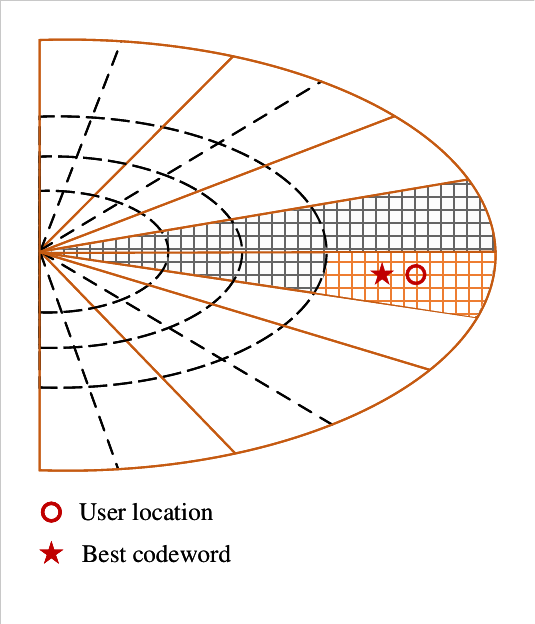}}
\caption{Beam training \cite{Weng_TVT_24}: b) joint angle and distance estimation.}\label{fig7}
\end{figure}

%%% Maybe needed; see later. %% 
%Note that the use of wide codeword can reduce the search of angle by a fraction. To further reduce this, tree-based angle alignment is introduced in [2], where multi-level search is conducted, with each lower level covers only the optimal .  
%%% Maybe needed; see later. %%

%Introduce grid-less solutions, and increase the number of sampling points. It may increase the complexity. Maybe used for the challenges;

\subsubsection{Conventional DFT-based beam training}
Due to the beam energy spread, conventional DFT-based codebooks cannot be directly applied for angle estimation in the near-field. However, recent studies have demonstrated that the received beam pattern from a conventional DFT-based codebook can still be leveraged for user angle estimation \cite{Zhang_WCL22, Zhou_TCOM24, Wu_TWC24}. In \cite{Zhang_WCL22}, a two-phase near-field beam training method was proposed to sequentially estimate the user's angle and range, thereby mitigating the need for a computationally expensive two-dimensional exhaustive search. Specifically, the conventional DFT-based codebook is initially employed for beam sweeping. As illustrated in Fig. 2, the true user angle is approximately centered within the angular support—defined as the region where the normalized beam gain exceeds a given threshold. Once the user angle is determined, the user range is estimated via a one-dimensional exhaustive search utilizing the polar-domain codebook. Notably, the beam sweeping in \cite{Zhang_WCL22} employs the entire antenna array, which can lead to excessive training overhead, especially in XL-MIMO systems where the number of antennas is substantial.  

To mitigate this overhead, a hybrid approach combining a sparse linear array and a central subarray was introduced in \cite{Zhou_TCOM24} for two-phase user angle identification. In the first phase, beam sweeping is performed using a DFT codebook constructed from a sparse linear array, and the angular support’s midpoint is designated as the user angle, following the approach in \cite{Zhang_WCL22}. However, the use of a sparse linear array introduces periodicity in the received beam pattern, leading to angular ambiguity. To resolve this issue, a second-phase beam sweeping over the central subarray is conducted to refine the angle estimate. Once the user angle is established, the polar-domain codebook is then employed to perform a one-dimensional search for range estimation, as in \cite{Zhang_WCL22}.

It is important to note that both \cite{Zhang_WCL22} and \cite{Zhou_TCOM24} utilize the polar-domain codebook for range estimation. As an on-grid beam training method, it only optimally aligns with users whose range coordinates coincide with predefined sampling points. However, since actual user positions are continuous and randomly distributed, any misalignment between the user’s true position and the nearest sampling grid results in beam training loss, thereby degrading overall system performance. A straightforward approach to improve beam training accuracy is to increase the density of range sampling points. However, this comes at the cost of increased training overhead.  
To address this limitation, an off-grid beam training method was proposed in \cite{Wu_TWC24}, which eliminates the need for a polar-domain codebook by relying solely on a conventional DFT-based codebook for joint angle and range estimation. In this method, the user angle is first identified using the DFT-based codebook, as in \cite{Zhang_WCL22}. The user range is then inferred based on the observation that the angular support width at a given user angle decreases as the user range increases, as depicted in Fig. 2 (blue and red solid curves). Since the exact functional relationship between angular support width and user range is difficult to derive analytically, a surrogate function is employed to estimate the user range in a semi-closed form.

%% new idea %%
% bisection method for find the user range, after identifying the angle; because of the monotonicity; or joint angle and distance evaluation; using only DFT codebook; 

%Although DFT-based solution. Although there exist beam split effect, DFT can still be used to obtain an approximate angle. Then, on this basis, estimate the distance or joint angle and distance. 

%Use central subarray with DFT for angle estimation and other solutions. Machine learning based on DFT as well. 

%Present a Table to compare the beam training overhead and performance of different solutions. 

%\subsubsection{Wideband Beam training} TTD. 

\begin{figure}
\centerline{\includegraphics[width=3.5in]{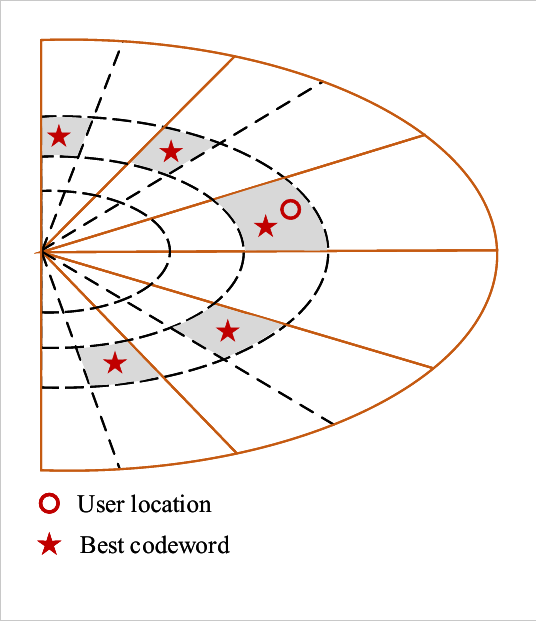}}
\caption{Beam training \cite{zhou2024}: a) multi-beam training.}\label{fig8}
\end{figure}

\section{Near-field Channel Estimation}
\subsection{Motivation}
The channel information obtained from beam training is inherently limited, as it primarily relies on signal strength feedback rather than full CSI. While this feedback is useful for beam alignment, it lacks the granularity required for optimal signal processing in XL-MIMO systems, particularly in multi-user scenarios. To fully harness the potential of XL-MIMO and enable efficient spatial multiplexing, comprehensive channel estimation is essential to acquire complete CSI.

In conventional massive MIMO systems, well-established channel estimation techniques, such as MMSE and LS, have been widely employed. These methods can still be applied to XL-MIMO systems operating in a TDD mode with fully digital precoding, where uplink-downlink channel reciprocity facilitates accurate CSI acquisition. However, the situation becomes considerably more complex in practical XL-MIMO implementations that rely on hybrid precoding-a widely adopted approach to reduce the number of required RF chains and associated hardware costs. Similarly, in FDD systems, where uplink and downlink channels are no longer reciprocal, downlink channel estimation faces additional challenges due to the increased feedback overhead required for CSI acquisition.

\begin{figure}
\centerline{\includegraphics[width=3.5in]{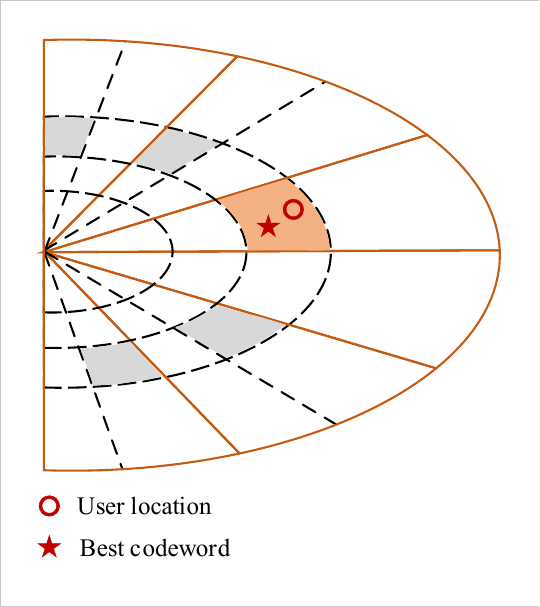}}
\caption{Beam training \cite{zhou2024}: b) remove angle ambiguity.}\label{fig9}
\end{figure}

\begin{figure}[ht!]
\centerline{\includegraphics[width=3.5in]{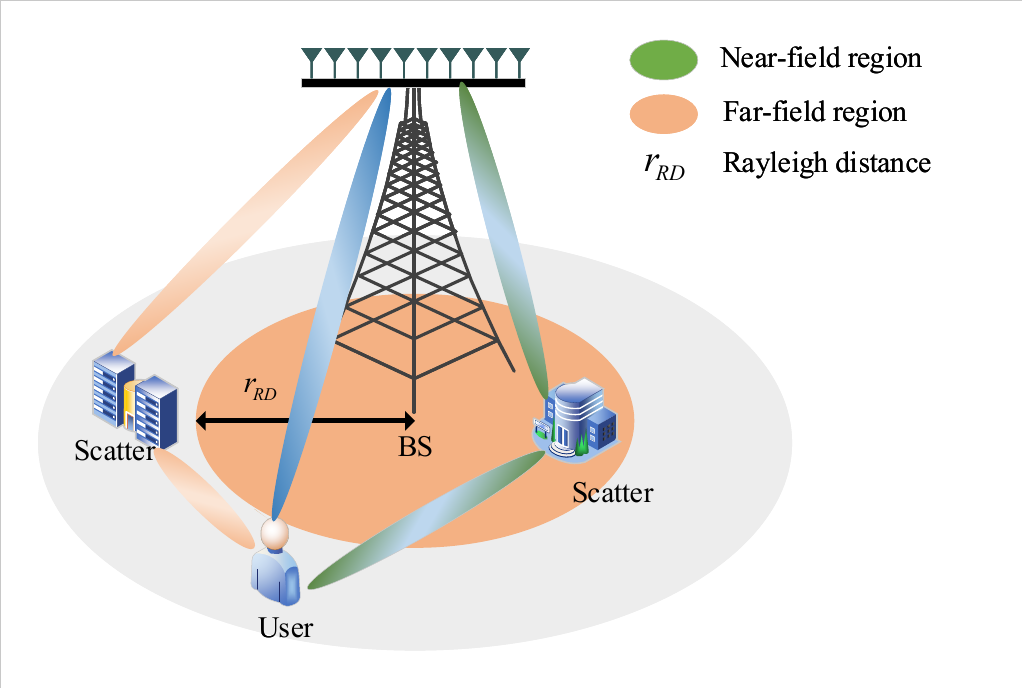}}
\caption{Systems with hybrid-field channels.} 
\label{fig:10}
\end{figure}

Traditional CS-based algorithms, which have been successfully applied in conventional massive MIMO, encounter limitations when extended to XL-MIMO systems. These algorithms typically exploit channel sparsity in the angular domain to reduce pilot overhead. However, this assumption of angular-domain sparsity no longer holds in the near-field propagation regime of XL-MIMO, where spherical wavefronts must be considered instead of the conventional planar wave approximation. In such scenarios, channel sparsity is distributed across both the angular and distance domains, necessitating novel estimation algorithms that jointly leverage these two dimensions.
Therefore, the development of advanced channel estimation techniques tailored for near-field XL-MIMO is crucial. These methods must account for the unique propagation characteristics of extra-large arrays while ensuring computational efficiency in hybrid precoding and FDD settings.

%Channel estimation can further improve the performance, especially under multi-user scenario, as it can better perform beamforming. In traditional solution, we exploit the spacial sparsity, especially under mmWave and Thz. More exactly, we perform a transform with the DFT matrix to make the channel matrix sparse. Then, we can apply other solutions like OMP for estimation. Once it is done, we reconstruct the channel. However, due to the existence of beam energy split, conventional DFT-based solution cannot be applied directly. 

\subsection{State-of-the-art}
\subsubsection{Channel Estimation in Uplink TDD Systems with Hybrid Beamforming Architecture}
%Similar to conventional massive MIMO systems, channel estimation for XL-MIMO systems operating in TDD mode should be performed in uplink, and then used for downlink transmission as well owing to channel reciprocity. However, in uplik TDD systems with hybrid beamforming architecture, the number of RF chains is much smaller than the number of antennas. Accordingly, the BS can no longer observe the received signals at each antenna simultaneously, thereby necessitating CS-based channel estimation to reduce the pilot overhead. However, existing far-field channel estimation schemes based on angular-domain sparsity can no longer be used due to the beam energy spread in near-field systems. To address this, polar-domain sparsity is exploited in \cite{Cui_TCOMM_22}, by taking into account both the angular and distance information. More exactly, by sampling at both the angular and distance domain, a transform matrix can be formed, with each column composed of its corresponding near-field steering vector at the sampling point. Then, the real user channel can be decomposed into the transform matrix multiplied by a sparse vector. On this basis, the sparse channel vector is estimated using the classical simultaneous orthogonal matching pursuit (SOMP) algorithm. To further improve the estimation accuracy, an off-grid polar-domain simultaneous iterative gridless weighted (P-SIGW) algorithm is proposed, which simultaneously recovers the path gains, angles, and distances. 

Similar to conventional massive MIMO systems, channel estimation for XL-MIMO systems operating in TDD mode should be performed in the uplink and subsequently leveraged for downlink transmission, exploiting channel reciprocity. However, in uplink TDD systems employing a hybrid beamforming architecture, the number of RF chains is significantly smaller than the total number of antennas. Consequently, the BS cannot directly observe the received signals at each antenna simultaneously, necessitating the use of CS-based channel estimation techniques to reduce pilot overhead.  
Despite their effectiveness in conventional massive MIMO, existing far-field channel estimation schemes that exploit sparsity in the angular domain become inapplicable in near-field XL-MIMO systems. This is due to the spatial beam energy spread caused by spherical wave propagation, which invalidates the traditional plane-wave assumption. To address this challenge, a polar-domain sparsity-based approach was proposed in \cite{Cui_TCOMM_22}, which jointly considers both angular and distance information for enhanced channel estimation.

Specifically, this approach samples the channel in both the angular and distance domains, constructing a transformation matrix whose columns correspond to near-field steering vectors at predefined sampling points. Under this formulation, the actual user channel can be expressed as the product of the transformation matrix and a sparse channel coefficient vector. Given this representation, the sparse channel vector is estimated using the classical simultaneous orthogonal matching pursuit (SOMP) algorithm.  
To further enhance estimation accuracy and mitigate the limitations of discretized sampling, an off-grid polar-domain simultaneous iterative gridless weighted (P-SIGW) algorithm is introduced. This algorithm jointly recovers key channel parameters, including path gains, angles, and distances, without being restricted to a fixed sampling grid. By iteratively refining parameter estimates, the P-SIGW method achieves superior accuracy in near-field CSI acquisition, making it a promising solution for hybrid precoded XL-MIMO systems.

While polar-domain-based transformation can sparsify the channel representation, it fails to fully exploit the inherent coupling between angular and distance parameters. Moreover, the two-dimensional dictionary used in such approaches introduces a significant storage burden and encounters challenges in balancing recovery accuracy and resolution. Specifically, according to CS theory, a larger sampling interval is desirable to reduce column coherence and improve recovery accuracy, whereas a smaller sampling interval is preferred for achieving finer resolution. This trade-off presents a fundamental limitation in polar-domain-based methods.

To overcome these challenges, a distance-parameterized angular-domain sparse representation was proposed in \cite{Zhang_TCOM23}. In this approach, the transformation matrix is constructed with a dimensionality dependent solely on angular resolution, significantly reducing storage burden. Additionally, the dictionary's coherence is greatly improved, as each column in the transformation matrix corresponds to a distinct angular component. 
Based on the proposed distance-parameterized angular-domain sparse representation, joint dictionary learning and sparse recovery techniques are developed for both LoS and multi-path propagation scenarios. Since distance is treated as a parameter in the transformation matrix, an iterative learning process is required to refine the dictionary and optimize sparse recovery.
Performance evaluations, conducted via normalized mean square error (NMSE) simulations, demonstrate that the proposed distance-parameterized angular-domain representation outperforms conventional polar-domain-based methods, particularly in multi-path environments. 
These findings highlight the advantages of incorporating a distance-adaptive dictionary structure, paving the way for more efficient and accurate channel estimation in near-field XL-MIMO systems.

It is important to note that in \cite{Cui_TCOMM_22} and \cite{Zhang_TCOM23}, the design of the analog beamformer is not explicitly considered. Instead, \cite{Cui_TCOMM_22} assumes an isotropic beamforming approach, while \cite{Zhang_TCOM23} employs a switching-based architecture. However, in hybrid beamforming systems, the BS can only observe signals after they have passed through the analog beamformer, making the choice of beamforming strategy a critical factor in the accuracy of channel estimation. Unlike beamformer design for data transmission, which can leverage available CSI, designing beamformers for channel estimation is significantly more challenging due to the lack of prior knowledge about the user's channel state.

To address this issue, a neural network-assisted joint optimization framework for beamformer design and localization was proposed in \cite{Jang_TCOM24}. This approach leverages deep learning to optimize the beamforming matrix in an environment-aware manner. During the training phase, the neural network takes the full CSI as input, followed by a weight matrix to approximate the beamformer. Once trained, during inference, these optimized weights are extracted to construct the beamformer, while the rest of the neural network functions as a localization module, processing received signals to estimate the user’s position. Following this initial localization, the estimated position parameters are further refined by exploiting range-dependent frequency selectivity characteristics. Presented results show that the proposed neural network-assisted beamformer design effectively adapts to specific propagation environments, leading to improvements in near-field channel estimation performance compared to conventional methods. 

Previous studies primarily focus on channel estimation for near-field propagation. However, as illustrated in Fig. 10, practical scenarios often involve hybrid-field propagation, where certain signal paths of a target user lie in the near-field while others extend into the far-field. In such hybrid-field environments, conventional sparsity assumptions—whether in the angular domain or the polar domain—no longer hold, necessitating the development of novel channel estimation strategies. 

To address this challenge, a hybrid-field channel estimation scheme was proposed in \cite{Yang_COMML23}. This approach employs a two-phase estimation process. In the first phase, a coarse estimation of far-field paths is conducted, leveraging traditional far-field codebooks. The second phase then refines the channel estimation by precisely identifying and estimating the near-field paths. Unlike existing hybrid-field estimation methods, which require prior knowledge of the ratio between far-field and near-field paths, the method in \cite{Yang_COMML23} adaptively determines this ratio by iteratively sweeping it in the second phase. The optimal ratio is identified by minimizing the residual error vector, as the dominant source of estimation error is attributed to noise.
Simulation results validate the effectiveness of the proposed method, demonstrating its superior performance compared to conventional techniques, particularly in hybrid-field channel conditions. 

\subsubsection{Channel Estimation in Downlink FDD Systems}
%In downlink FDD systems, each user performs channel estimation independently from others using the received pilot signal from the BS. Often, it is assumed that the user is equipped with single antenna. In this case, the signal observed at the user is only a single complex value, i.e., the product of the precoded pilot with the channel plus noise. Note that in uplink TDD system with hybrid beamforming architecture, the size of observed signal at the BS equals to the number of RF chains, which is often larger than 1. Therefore, if we compare the two systems, it is clear that a larger number of pilots is needed for channel estimation in downlink FDD systems to achieve the same estimation accuracy than in uplink TDD system. Indeed, to use the conventional channel estimation methods such as LS or MMSE, the number of pilot symbols should be no less than the number of antennas, which can be extremely large in XL-MIMO systems. To reduce the pilot overhead, CS-based methods should be developed by exploiting the sparse nature of the channel. Nonetheless, multiple pilot symbols are still needed to obtain a decent estimation accuracy in downlink FDD systems. In this case, the received signal vector after a simple conjugation has exactly the same form with its counterpart in uplink TDD systems. Therefore, the channel estimation methods proposed for uplink TDD system can be directly applied to downlink FDD systems and vice versa.  

In downlink FDD systems, each user independently performs channel estimation using the received pilot signal from the BS. It is often assumed that the user is equipped with a single antenna, in which case the observed signal is a single complex value—representing the product of the precoded pilot and the channel, plus noise. In contrast, in uplink TDD systems with a hybrid beamforming architecture, the dimensionality of the observed signal at the BS corresponds to the number of RF chains, which is typically greater than one. As a result, downlink FDD systems require a larger number of pilot symbols to achieve the same channel estimation accuracy as uplink TDD systems. In fact, to use conventional channel estimation methods such as LS or MMSE, the number of pilot symbols must be at least equal to the number of antennas—an impractical requirement in XL-MIMO systems, where the antenna count can be extremely high.

To mitigate pilot overhead, CS-based methods can be employed by exploiting the sparse nature of the channel. However, multiple pilot symbols are still necessary to achieve a reasonable estimation accuracy in downlink FDD systems. Notably, after a simple conjugation, the received signal vector in a downlink FDD system takes exactly the same mathematical form as its counterpart in an uplink TDD system. Consequently, channel estimation methods developed for uplink TDD systems can be directly applied to downlink FDD systems, and vice versa.

In \cite{wang2024}, the channel estimation problem for a downlink THz system was studied. To reduce the storage burden and mutual coherence in the polar-domain channel representation, a novel dictionary was proposed. This dictionary was constructed by multiplying the conventional DFT matrix with a diagonal matrix that incorporates near-field properties. It was shown that, in the near-field regime, this new dictionary leads to a sparser channel representation.
Building on this foundation, pattern-coupled sparse Bayesian learning—a block-sparse signal recovery algorithm that does not require prior knowledge of block partitioning—was employed to accurately estimate the near-field channel. Additionally, by leveraging the block-sparse structure of the channel representation, the proposed method significantly reduced training overhead. Numerical results demonstrated that the proposed approach achieved lower channel estimation error compared to conventional polar-domain-based methods. 
%In \cite{wang2024}, the channel estimation problem for a downlink THz system was investigated. To reduce the storage burden and mutual coherence of polar-domain channel representation, a new dictionary was proposed, by multiplying the DFT matrix with a diagonal matrix that incorporate the near-field properties. It was shown that the new dictionary could make the channel representation sparser, when the channel is in the near-field. On this basis, the pattern-coupled sparse Bayesian learning, a block-sparse signal recovery algorithm that requires no prior knowledge of block partition knowledge of block partition was employed to achieve an accurate estimate of the near-field channel. It was also shown that by exploiting the block-sparse structure of the channel representation, the proposed method could greatly reduce the training overhead. Numerical results showed that the proposed method obtains a lower channel estimation error when compared with the polar-domain-based ones. 

\subsubsection{Channel Estimation for Point-to-Point XL-MIMO System} 
%In point-to-point XL-MIMO system, both the transmitter and receiver are equipped with a massive number of antennas. Meanwhile, to reduce the power consumption and hardware cost, the hybrid beamforming architecture is often adopted at the transceivers. Previously, for both the LoS and NLoS paths of such systems, the MIMO channel was modeled using two MISO channels. However, \cite{Lu_TCOM23} points out that such a modeling is not accurate for the LoS path component, since each transmitter-receiver antenna pair experiences different propagation paths as illustrated in Fig. 6. Instead, the geometric free space propagation modeling should be adopted for each transmitter-receiver antenna pair. Accordingly, the LoS path component cannot be presented by polar-domain channel with transformation matrices. To address the channel estimation in such a mixed LoS/NLoS scenario, a two stage algorithm was proposed. In the firs stage, the LoS path component is first estmated by searching collection with coarse on-grid parameters, and then refined by iteration optimization. On this basis, the NLoS path components are estimated using OMP-based estimation with their polar-domain sparsity. Simulation results showed that the proposed two stage channel estimation scheme achieved better NMSE performance than near-field codebook based channel estmation schemes.   

In a point-to-point XL-MIMO system, both the transmitter and receiver are equipped with a massive number of antennas. 
%To reduce power consumption and hardware costs, a hybrid beamforming architecture is often adopted at the transceivers. 
Traditionally, the MIMO channel in such systems has been modeled as two separate MISO channels for both the LoS and non-line-of-sight (NLoS) components. However, as noted in \cite{Lu_TCOM23}, this modeling approach is inaccurate for the LoS component, as each transmitter-receiver antenna pair experiences distinct propagation paths, as illustrated in Fig. 11. Instead, the geometric free-space propagation model should be applied to each transmitter-receiver antenna pair individually. Consequently, the LoS component cannot be accurately represented using a polar-domain channel model with transformation matrices.

To address channel estimation in such a mixed LoS/NLoS scenario, a two-stage algorithm was proposed. In the first stage, the LoS component is estimated by searching a collection of coarse on-grid parameters, which are then refined through iterative optimization. Based on this initial estimate, the NLoS components are subsequently estimated using an OMP-based approach that exploits their polar-domain sparsity. Simulation results demonstrated that the proposed two-stage channel estimation scheme outperforms near-field codebook-based methods in terms of NMSE.

The LoS modeling and estimation method described above was adopted in \cite{Ruan_COMML24}, where a multidimensional OMP with parameter refinement algorithm was proposed to estimate the NLoS path components with reduced complexity. The key idea is to represent the sparse channel using a set of independent lower-dimensional polar-domain dictionaries rather than a single high-dimensional one. Numerical results demonstrated that the proposed algorithm achieves a better balance between NMSE performance and computational complexity compared to existing CS-based algorithms.

\begin{figure}[ht!]
\centerline{\includegraphics[width=3.5in]{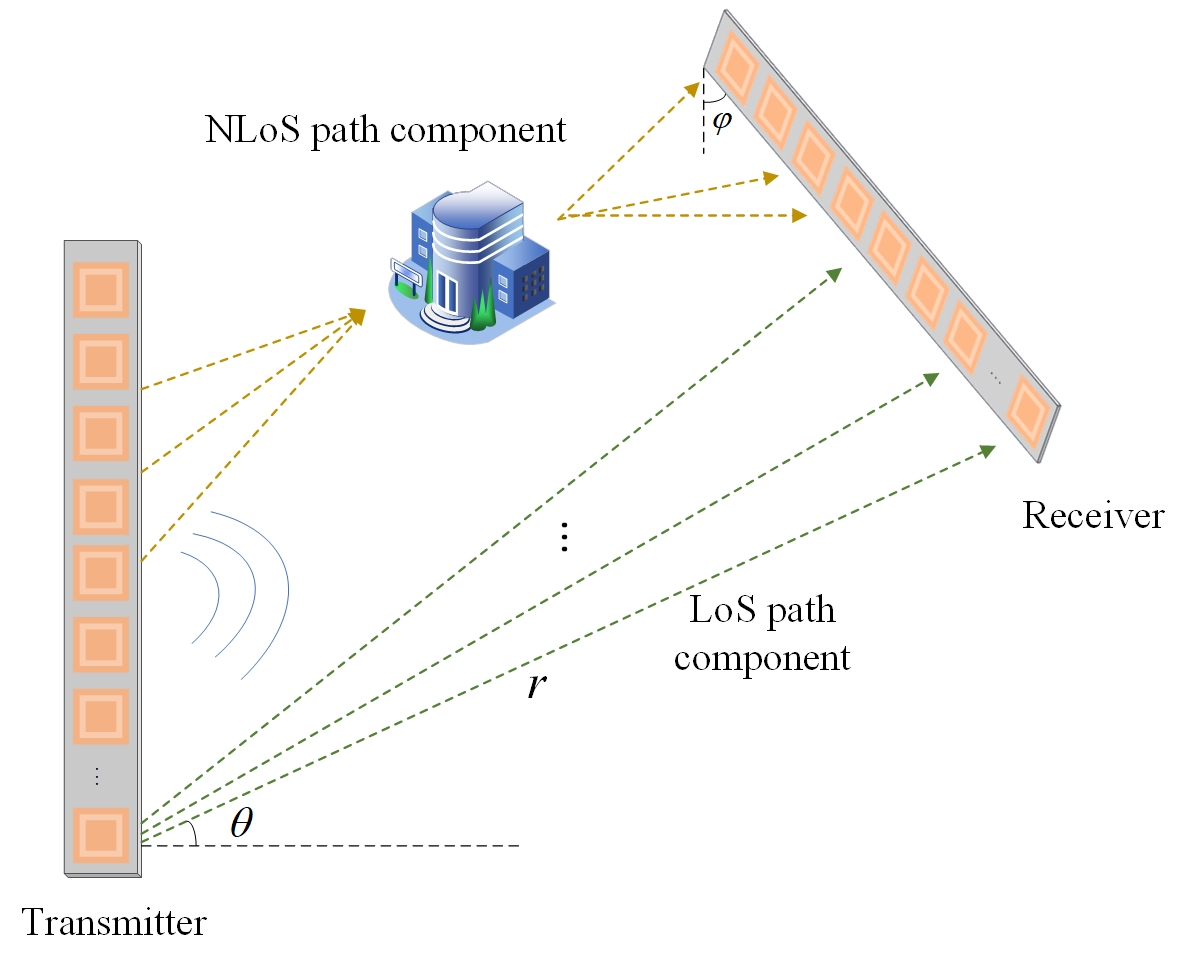}}
\caption{Point-to-point XL-MIMO systems.} 
\label{fig:Low_2}
\end{figure}

Unlike \cite{Lu_TCOM23} and \cite{Ruan_COMML24}, which model the LoS and NLoS path components separately, \cite{Shi_TCOM24} introduces a unified channel model that characterizes both LoS and NLoS paths in a consistent framework. Specifically, the LoS path is represented as two MISO channels multiplied elementwise by a Vandermonde phase matrix, resulting in a structure analogous to its NLoS counterpart. Building on this model, a unified LoS/NLoS orthogonal matching pursuit algorithm is proposed for channel estimation, which is further enhanced through a three-stage multiple-measurement-vector approach to reduce computational complexity. Numerical results demonstrate the effectiveness of this method in sparse channel estimation.

%\subsection{Proposed solution} Observation: beam training solution can often be applied for channel estimation. We exploit this concept by using DFT to estimate angle and the distance search over the effective Rayleigh distance. Compare with existing solutions. 

%\begin{figure}[ht!]
%\centering
%\includegraphics[width=1\linewidth]{Images/fig1a.png}
%\caption{Achievable max-min rate versus the transceiver hardware impairment factor.} 
%\label{fig:HWI_1}
%\end{figure}

\section{Challenges and Open Directions}
\subsection{Real Measurement Data-Based Performance Evaluation}
In most existing studies, the performance of beam training and channel estimation is evaluated using artificially generated channels based on the near-field channel model. However, in practical scenarios, real-world channel characteristics may deviate from those predicted by the assumed near-field model due to factors such as hardware impairments, environmental variations, and multipath effects. Consequently, the effectiveness of the proposed beam training and channel estimation techniques under real-world conditions remains uncertain. To bridge this gap, it is crucial to assess their performance using real-world measured channel data, ensuring their robustness and practicality in actual deployment scenarios \cite{Tang24}.

\subsection{Sensing-Enhanced Beam Training and Channel Estimation}
Due to the spherical wave propagation characteristics in near-field communications, existing beam training and channel estimation techniques must consider both the angular and range domains. This requirement leads to a large dictionary and relatively high computational complexity. Leveraging sensing capabilities to acquire localization information for both users and scatterers presents a promising approach to improving beam training and channel estimation performance while reducing the codebook size and computational burden.
For instance, in \cite{Liu_JSAC25}, power sensors are embedded into the antenna elements to first estimate the user’s location. Based on this information, a more compact and efficient dictionary is constructed, facilitating lightweight near-field channel estimation. Numerical results demonstrate that the proposed method enhances channel estimation accuracy while reducing pilot overhead. 

\subsection{Beam Training and Channel Estimation for FR3 Band}
Frequency Range 3 (FR3), spanning approximately 7–24 GHz, is anticipated to be one of the most critical new spectrum bands for 6G. Its propagation characteristics differ significantly from those observed in both mmWave)/THz bands and sub-6 GHz frequencies. However, research on beam training and channel estimation for the FR3 band remains in its early stages, requiring substantial investigation.
A key challenge in this context is determining whether FR3 channels will primarily exhibit near-field characteristics or if they will operate in a hybrid regime, incorporating both near-field and far-field components. Additionally, given the broad frequency range of FR3, its lower and upper portions may exhibit distinct propagation behaviors, further complicating channel modeling and system design. Therefore, a deeper understanding of FR3 channel spatial properties and their impact on beam training and channel estimation is essential for optimizing future 6G communication systems \cite{xu2025}.

{\color{black}
\subsection{Machine Learning and AI-Assisted Beam Training and Channel Estimation}
Another promising research direction is the integration of machine learning and artificial intelligence (AI) techniques into beam training and channel estimation for XL-MIMO systems. Traditional approaches rely heavily on analytical channel models and handcrafted algorithms, which may not fully capture the complex propagation characteristics in near-field environments. Data-driven methods can learn channel representations and beamforming strategies directly from observations, enabling more efficient beam prediction, channel parameter estimation, and codebook design. By leveraging additional side information such as user location, sensing data, or historical channel measurements, learning-based approaches may further reduce training overhead and improve estimation accuracy. However, important challenges remain, including dataset requirements, generalization across environments, and the design of lightweight models suitable for practical deployment.
}

\section{Conclusion} 
\label{Sec:Conclusion}
XL-MIMO is poised to play a pivotal role in next-generation wireless communication systems, offering unprecedented improvements in spectral efficiency and spatial resolution. However, the transition from conventional massive MIMO to XL-MIMO introduces new challenges, particularly in beam training and channel estimation, due to the shift from planar to spherical wave propagation.  
In this article, we provided a structured and in-depth review of the state-of-the-art beam training and channel estimation methods, categorizing existing approaches and analyzing their effectiveness in near-field propagation environments. 
Looking ahead, future research efforts should focus on developing adaptive and scalable solutions, potentially leveraging sensing-aided techniques to enhance system performance. Moreover, experimental validation using real-world channel measurements will be essential to ensure practical applicability. {\color{black}In particular, integrating machine learning techniques with model-based signal processing may provide powerful tools for improving beam training efficiency and channel estimation accuracy in near-field XL-MIMO systems.}

\bibliographystyle{IEEEtran}
\bibliography{biblio}

\end{document}